# Missing Member of the $J^{PC} = 2^{--}$ Nonet in Extreme Conditions

J. Y. Süngü[1,a)], A. Türkan[2], E. Sertbakan[1] and E. Veli Veliev[1]

[1]*Department of Physics, Kocaeli University, 41001 Izmit, Turkey*
[2]*Özyeğin University, Department of Natural and Mathematical Sciences, Çekmeköy, Istanbul, Turkey*

a)Corresponding author:jyilmazkaya@kocaeli.edu.tr

**Abstract.** The isoscalar member of the axial-tensor family that should take place in the further states of the PDG is still absent. In this study, analyzes are made as to whether one of the X states whose quantum numbers are unknown in this group, is a candidate for the missing resonance under the assignment $2^3D_2$. Moreover, replacing the time evolution operator with the thermal average one, we construct the modified correlator satisfying Thermal QCD sum rules approach. We determine that the hadronic parameters of the considered nonet member are sensitive to the increment of temperature. This knowledge can be helpful to complete the $J^{PC} = 2^{--}$ light meson nonet and also explore the hot medium behaviors at upcoming heavy-ion collision experiments.

## INTRODUCTION

During the first few seconds after the Big Bang, early universe is thought to be filled with quark-gluon plasma (QGP), rather than hadrons [1]. Other possible location at which we may find this new form of matter is in the interior of neutron stars. In hot medium, hadronic quantities shows a strong drop in a certain temperature range ($\sim T_c = 155$ MeV) [2], indicating deconfinement and restoration of chiral symmetry [3].

Since the chiral symmetry of QCD Lagrangian is almost exact in the light quark sector, the temperature behavior of their properties provide information on the chiral symmetry restoration. Looking at the light hadron spectrum there is one more space to be filled in the $J^{PC} = 2^{--}$ nonet with $I_3 = 0$. This is our another motivation for studying the yet-unobserved meson called $\phi_2$. Moreover PDG and Regge Trajectory model data [4, 5] predict different masses for this axial-tensor family members. It is necessary to clear the case to determine and fix the masses for the ground states belonging to light unflavored nonet. $\phi_2$ resonance is the mixture of the SU(3) wave function $\psi_8$ and $\psi_1$, namely $\phi_2 = \psi_8 cos\theta - \psi_1 sin\theta$, where $\theta$ is the mixing angle of the $J^{PC} = 2^{--}$ nonet. The physical $\phi_2$ state is the linear combination of the SU(3) octet and singlet states: $\psi_8 = 1/\sqrt{6}(u\bar{u} + d\bar{d} - 2s\bar{s})$ and $\psi_1 = 1/\sqrt{3}(u\bar{u} + d\bar{d} + s\bar{s})$. The mixing angle of singlet and octet states can be neglected because of the relatively small effect. Therefore $\phi_2$ state is considered as a pure octet state.

In this paper, including the finite-temperature effects in the QCD sum rules theory, the thermal behavior of the parameters of the axial-tensor $\phi_2$ resonance is investigated up to dimension five taking into account quark and quark-gluon mixed condensates. The outline of the paper can be summarized as follows. In section 2 we give a brief review of the Thermal QCD sum rules (TQCDSR) model for $\phi_2$ state. Section 3 deals with some numeric technicalities, namely determination of the nonperturbative contributions with thermal settings.

## THERMAL QCD SUM RULES for the $\phi_2$ RESONANCE

TQCDSR is one of the most effective technique that can be used in order to study the nonperturbative regime of strong interactions [6]. This approach is based on the assumptions that quark-hadron duality and operator product expansions (OPE) are valid at finite temperatures. In order to find the mass and decay constant of the $\phi_2$ resonance by the TQCDSR model, first a two-point temperature dependent correlation function is identified with;

$$\Pi_{\mu\nu,\alpha\beta}(p,T) = i \int d^4x e^{ip\cdot(x-y)} Tr\left\{\varrho\mathcal{T}\left[J_{\mu\nu}(x)J^{\dagger}_{\alpha\beta}(y)\right]\right\}_{y\to 0}, \quad (1)$$

where $J_{\mu\nu}$ is the interpolating current for $\phi_2$ meson, $\mathcal{T}$ is the time ordering product. The thermal density matrix can be defined as $\varrho = e^{-H/T}/Tr(e^{-H/T})$ here $T$ is the temperature of the medium and $H$ is the QCD Hamiltonian. The interpolating current of $\phi_2$ state is selected as in the following form [7]

$$J^{\phi_2}_{\mu\nu}(x) = \frac{1}{2\sqrt{6}}\left\{\left[\bar{u}(x)\gamma_\mu\gamma_5 \overleftrightarrow{\mathcal{D}}_\nu(x)u(x) + \bar{d}(x)\gamma_\mu\gamma_5 \overleftrightarrow{\mathcal{D}}_\nu(x)d(x) - 2\bar{s}(x)\gamma_\mu\gamma_5 \overleftrightarrow{\mathcal{D}}_\nu(x)s(x)\right] + [\mu \leftrightarrow \nu]\right\}, \quad (2)$$

where $\overleftrightarrow{\mathcal{D}}_\mu(x)$ shows the covariant derivative with respect to four-$x$ acting on both left and right side. There are two parts we need to calculate: First one is the "*phenomenological side*" of the correlation function.

• **Phenomenological Part** : At this side, a complete set of intermediate phenomenological states with the same quantum numbers are sandwiched into Eq. (1), then integrals over four-$x$ are carried out. The correlation function can be defined by the matrix elements of interpolating current

$$\Pi^{phen}_{\mu\nu,\alpha\beta}(p,T) = \frac{\langle\Omega|J_{\mu\nu}(0)|\phi_2\rangle\langle\phi_2|\bar{J}_{\alpha\beta}(0)|\Omega\rangle}{m^2_{\phi_2}(T) - p^2} + ..., \quad (3)$$

here $\Omega$ indicates the hot medium, dots symbolize the contributions originating from the excited states and continuum. The matrix elements $\langle\Omega|J_{\mu\nu}(0)|\phi_2\rangle$ and $\langle\phi_2|\bar{J}_{\alpha\beta}(0)|\Omega\rangle$ are described depending on the decay constant $f_{\phi_2}$ and the mass $m_{\phi_2}$ in the following form

$$\langle\Omega|J_{\mu\nu}(0)|\phi_2\rangle = f_{\phi_2}(T)m^3_{\phi_2}(T)\,\varepsilon_{\mu\nu}, \quad \langle\phi_2|\bar{J}_{\alpha\beta}(0)|\Omega\rangle = f_{\phi_2}(T)m^3_{\phi_2}(T)\,\varepsilon'_{\mu\nu}, \quad (4)$$

here $\varepsilon_{\mu\nu}$ represents the polarization tensor and the below relation is valid:

$$\varepsilon_{\mu\nu}\varepsilon'_{\alpha\beta} = \frac{1}{2}\eta_{\mu\alpha}\eta_{\nu\beta} + \frac{1}{2}\eta_{\mu\beta}\eta_{\nu\alpha} - \frac{1}{3}\eta_{\mu\nu}\eta_{\alpha\beta}, \text{ here } \eta_{\mu\nu} = -g_{\mu\nu} + \frac{p_\mu p_\nu}{m^2_{\phi_2}}. \quad (5)$$

By placing Eqs. (4-5) into Eq. (3), the last expression of the correlation function for the phenomenological part is:

$$\Pi^{phen}_{\mu\nu,\alpha\beta}(p,T) = \frac{f^2_{\phi_2}(T)m^6_{\phi_2}(T)}{m^2_{\phi_2}(T) - p^2}\left[\frac{1}{2}(g_{\mu\alpha}g_{\nu\beta} + g_{\mu\beta}g_{\nu\alpha})\right] + other\ structures. \quad (6)$$

• **Theoretical Part**: Second step is to compute the correlation function putting the interpolating currents in Eq. (2) into Eq. (1). After some straightforward computations we get the theoretical side of the correlation function as below:

$$\Pi^{\phi_2}_{\mu\nu,\alpha\beta}(p,T) = \frac{3i}{32}\int d^4x\, e^{ip\cdot(x-y)}\Bigg\{Tr\bigg[\bigg(-\overrightarrow{\mathcal{D}}_\beta(y)S_u(y-x)\gamma_\mu\gamma_5\overrightarrow{\mathcal{D}}_\nu(x)S_u(x-y)\gamma_\alpha\gamma_5 + S_u(y-x)\gamma_\mu$$
$$\times\gamma_5\overrightarrow{\mathcal{D}}_\nu(x)\overrightarrow{\mathcal{D}}_\beta(y)S_u(x-y)\gamma_\alpha\gamma_5 + \overrightarrow{\mathcal{D}}_\beta(y)\overrightarrow{\mathcal{D}}_\nu(x)S_u(y-x)\gamma_\mu\gamma_5 S_u(x-y)\gamma_\alpha\gamma_5 - \overrightarrow{\mathcal{D}}_\nu(x)S_u(y-x)\gamma_\mu\gamma_5$$
$$\times\overrightarrow{\mathcal{D}}_\beta(y)S_u(x-y)\gamma_\alpha\gamma_5\bigg) + (\beta\leftrightarrow\alpha) + (\nu\leftrightarrow\mu) + (\beta\leftrightarrow\alpha,\nu\leftrightarrow\mu)\bigg] + [u\to d] + 4[u\to s]\Bigg\}_{y\to 0}. \quad (7)$$

Inserting the thermal light quark propagator $S_q(x-y)$ in Eq. (7) defined as in the following form [8]:

$$S^{ij}_q(x-y) = i\frac{\slashed{x}-\slashed{y}}{2\pi^2(x-y)^4}\delta_{ij} - \frac{m_q}{4\pi^2(x-y)^2}\delta_{ij} - \frac{\langle\bar{q}q\rangle_T}{12}\delta_{ij} - \frac{(x-y)^2}{192}m^2_0\langle\bar{q}q\rangle_T\left[1 - i\frac{m_q}{6}(\slashed{x}-\slashed{y})\right]\delta_{ij}$$
$$+ \frac{i}{3}\bigg[(\slashed{x}-\slashed{y})\bigg(\frac{m_q}{16}\langle\bar{q}q\rangle_T - \frac{1}{12}\langle u^\mu\Theta^f_{\mu\nu}u^\nu\rangle\bigg) + \frac{1}{3}\bigg(u\cdot(x-y)\slashed{u}\langle u^\mu\Theta^f_{\mu\nu}u^\nu\rangle\bigg)\bigg]\delta_{ij}$$
$$- \frac{ig_s G_{\mu\nu}}{32\pi^2(x-y)^2}\bigg((\slashed{x}-\slashed{y})\sigma^{\mu\nu} + \sigma^{\mu\nu}(\slashed{x}-\slashed{y})\bigg)\delta_{ij}, \quad (8)$$

here $\Theta^f_{\mu\nu}$ and $u_\mu$ are the fermionic part of the energy momentum tensor and the four-velocity of the hot medium, respectively. The quark condensates depending on temperature are governed by the vacuum condensates in the rest frame $u_\mu = (1,0,0,0)$, $u^2 = 1$ [9]. After standard manipulations, the correlation function in terms of the chosen Lorentz structures is written as:

$$\Pi^{\text{theo}}_{\mu\nu,\alpha\beta}(p^2, T) = \Pi^{\text{theo}}(p^2, T)\left\{\frac{1}{2}(g_{\mu\alpha}g_{\nu\beta} + g_{\mu\beta}g_{\nu\alpha})\right\} + \text{other structures.} \qquad (9)$$

Scalar part of the correlation function in Eq. (9) can be divided into two part as perturbative $\Gamma(p^2, T)$ and nonperturbative part $\widetilde{\Gamma}(p^2, T)$, respectively; $\Pi^{\text{theo}}(p^2, T) = \Gamma(p^2, T) + \widetilde{\Gamma}(p^2, T)$. The analyticity of the function $\Pi(p^2)$ in the whole imaginary plane of the variable $q^2$ except the positive part of the real axis allows one write down the dispersion relation with a spectral function $\rho(s) = Im[\Gamma(s, T)]/\pi$:

$$\Gamma(p, T) = \int \frac{\rho(s)}{s - p^2}\, ds + \text{subtracted terms.} \qquad (10)$$

Anymore we have the correlators for both the phenomenological and theoretical sides differentiating the terms with regard to their structures. To eliminate the highest orders from the ground state we take derivative of $p^2$ in both side of the correlator based on the philosophy of the QCDSR and assuming the quark-hadron duality, then we obtain the ground-state decay constant sum rule for $\phi_2$ as

$$f^2_{\phi_2}(T) = m^{-6}_{\phi_2}(T)\, e^{m^2_{\phi_2}/M^2}\left[\int_{s_{min}}^{s_0(T)} ds\, \rho^{\text{pert}}(s)\, e^{-s/M^2} + \widehat{\mathcal{B}\widetilde{\Gamma}}(p^2, T)\right], \qquad (11)$$

where $\hat{\mathcal{B}}$ represent the Borel transformation. So the mass sum rule is derived from the Eq. (11) easily by taking derivative in terms of $(-1/M^2)$ where $M^2$ is the Borel mass parameter. Next the mass sum rule for the ground-state $\phi_2$ meson is obtained as:

$$m^2_{\phi_2}(T) = \frac{\int_{s_{min}}^{s_0(T)} ds\, \rho^{\text{pert}}(s)\, s\, e^{-s/M^2} + \frac{d}{d(-1/M^2)}\widehat{\mathcal{B}\widetilde{\Gamma}}(p^2, T)}{\int_{s_{min}}^{s_0(T)} ds\, \rho^{\text{pert}}(s)\, e^{-s/M^2} + \widehat{\mathcal{B}\widetilde{\Gamma}}(p^2, T)}, \qquad (12)$$

here $s_{min}$ is the square of the sum of the quark contents of the $\phi_2$, $s_0(T)$ is the thermal continuum threshold, which separates the contribution of $\phi_2$ from the "higher resonances and continuum". As a result of our evaluations, the spectral density function and the non-perturbative contribution to the correlation function are reached as follows:

$$\rho_{\phi_2} = \frac{12s^2 - 5s(m^2_d + m^2_u + 4m^2_s)}{320\pi^2}, \qquad (13)$$

$$\widetilde{\Gamma}_{\phi_2}(p, T) = \frac{8\langle u\Theta^f u\rangle(p \cdot u)^2}{27p^2} - \frac{41m^2_0(m_d\langle \bar{d}d\rangle + m_u\langle \bar{u}u\rangle + 4m_s\langle \bar{s}s\rangle)}{288p^2}. \qquad (14)$$

## NUMERIC CONSEQUENCES

Now the input parameters used in our computations to analyze the obtained sum rules in Eqs. (11-12) will be given here. For the quark and mixed condensates we employ $\langle \bar{q}g_s\sigma Gq\rangle = m^2_0\langle \bar{q}q\rangle$, where $m^2_0 = (0.8 \pm 0.2)$ GeV$^2$, $\langle 0|\bar{u}u|0\rangle = \langle 0|\bar{d}d|0\rangle = -(0.24 \pm 0.01)^3$ GeV$^3$, $\langle 0|\bar{s}s|0\rangle = -0.8(0.24 \pm 0.01)^3$ GeV$^3$ [10, 11]. The masses of $u$, $d$ and $s$ quarks; $m_u = (2.16^{+0.49}_{-0.26})$ MeV, $m_d = (4.67^{+0.48}_{-0.17})$ MeV and $m_s = (93^{+11}_{-5})$ MeV [4]. The normalized thermal quark condensate is used fitting Lattice data from Ref. [12, 13] and the fermionic part of the energy density is parameterized as follows [14]:

$$\frac{\langle \bar{q}q\rangle_T}{\langle 0|\bar{q}q|0\rangle} = \lambda_1 e^{aT} + \lambda_2, \quad \frac{\langle \bar{s}s\rangle_T}{\langle 0|\bar{s}s|0\rangle} = \lambda_3 e^{bT} + \lambda_4, \quad \langle u^\mu \Theta^f_{\mu\nu} u^\nu\rangle_T = T^4 e^{(\lambda_5 T^2 - \lambda_6 T)} - \lambda_7 T^5, \qquad (15)$$

here $a = 0.040$ MeV$^{-1}$, $b = 0.516$ MeV$^{-1}$, $\lambda_1 = -6.534 \times 10^{-4}$, $\lambda_2 = 1.015$, $\lambda_3 = -2.169 \times 10^{-5}$, $\lambda_4 = 1.002$, $\lambda_5 = 113.867$ GeV$^{-2}$, $\lambda_6 = 12.190$ GeV$^{-1}$, $\lambda_7 = 10.141$ GeV$^{-1}$ and $q = u$ or $d$ quarks.

In Eqs. (11) and (12), the mass and decay constant sum rules rely on the Borel mass parameter. Thus we fix the interval of the Borel window $M^2$ and continuum threshold $s_0$, so that our results are almost insensitive to their variations since they are not completely physical quantities. Taking into account the above criteria and settling the Borel window $1.9 \text{ GeV}^2 \leq M^2 \leq 2.1 \text{ GeV}^2$ and the continuum threshold $s_0 = 5.52 \text{ GeV}^2$ for the $\phi_2$ resonance, we found the following results at vacuum:

$$m_{\phi_2} = 1843 \text{ MeV}, \quad f_{\phi_2} = 6.67 \times 10^{-2}.$$

Afterwards we used the following fit functions to parameterize the temperature dependence of the mass and decay constant:

$$m_{\phi_2}(T) = a + b\left(\frac{T}{T_c}\right)^\alpha, \quad f_{\phi_2}(T) = c + d\left(\frac{T}{T_c}\right)^\beta \tag{16}$$

here $T_c$ is the critical temperature and the variables $a, b, c, d, \alpha$ and $\beta$ are presented in Table 1.

**TABLE 1.** Values of fit parameters $a, b, \alpha, c, d$ and $\beta$.

| $a$ [GeV] | $b$ [GeV] | $\alpha$ | $c$ | $d$ | $\beta$ |
|---|---|---|---|---|---|
| 1.857 | −0.189 | 6.652 | $6.628 \times 10^{-2}$ | $-1.349 \times 10^{-3}$ | 8.697 |

Additionally we worry about whether $\phi_2$ meson is made up from only "$s\bar{s}$" content which is a probability according to PDG. Then we followed the same procedure aforementioned and get the following results at $T = 0$:

$$m_{\phi_2} = 1837 \text{ MeV}, \quad f_{\phi_2} = 6.72 \times 10^{-2}.$$

Mass value of $\phi_2$ resonance is predicted as 1904 MeV in terms of Regge Trajectory model [5]. Also very recently Abreu and et al. estimate the mass of $\phi_2$ as $m_{\phi_2} = 1850$ MeV employing the Coulomb gauge Hamiltonian approach to QCD [15]. These values are compatible with our results of the calculations made with both "$s\bar{s}$" or "$u\bar{u} + d\bar{d} + s\bar{s}$" combinations for $\phi_2$ state [16]. Further when we compare the decay constant values for $\phi_2$ with Ref. [7], the findings are in agreement with each other.

As a result, at $T = 0$ limit our mass value is very well consistent with the results in Refs. [5, 15, 17]. On the other hand as seen from our analyzes $X(2020)$, $X(2075)$ and $X(2080)$ states can not be the candidates for the wanted member of the $J^{PC} = 2^{--}$ nonet. Also we investigate the temperature dependence of the missing axial-tensor member with quantum number $J^{PC} = 2^{--}$ for "$u\bar{u} + d\bar{d} + s\bar{s}$" combination constructed from the TQCDSR using the first five dimensional terms, together with modified thermal hadronic spectral function. Looking at the mass and decay constant of $\phi_2$ resonance in hot medium they start to decrease at $T/T_c \simeq 0.7$ and change $\sim 10.14\%, 2\%$ respectively with rising temperature up to the critical temperature. It is of great importance both for theorist and the heavy-ion community that thermal effects be understood, as they are an essential ingredient in establishing the observation of the QGP.